\documentclass[showpacs,notitlepage,prd,aps,longbibliography,twocolumn]{revtex4-1}
\usepackage{graphicx}\usepackage{amsmath}\usepackage{slashed}
\newcommand{\be}{\begin{equation}}\newcommand{\ee}{\end{equation}}
\newcommand{\bea}{\begin{eqnarray}}\newcommand{\eea}{\end{eqnarray}}
\newcommand{\nn}{\nonumber}
\newcommand{\pa}{\partial}

\newcommand{\al}{\alpha}
\newcommand{\ep}{\varepsilon}
\newcommand{\om}{\omega}\newcommand{\Om}{\Omega}
\renewcommand{\phi}{\varphi}
\renewcommand{\kappa}{\varkappa}
\newcommand{\q}{q_{\rm B}}
\newcommand{\x}{\mathbf{x}}
\newcommand{\s}{\mathbf{s}}
\newcommand{\p}{\mathbf{p}}
\newcommand{\n}{\mathbf{n}}\newcommand{\m}{\mathbf{m}}
\renewcommand{\k}{\mathbf{k}}
\renewcommand{\j}{\mathbf{j}}
\newcommand{\A}{\mathbf{A}}
\newcommand{\B}{\mathbf{B}}\newcommand{\E}{\mathbf{E}}

\newcommand{\bxi}{\boldsymbol{\xi}}
\newcommand{\Ref}[1]{(\ref{#1})}

\newcommand{\elm}{electromagnetic~}
\begin{document}
\title{Monoatomically thin polarizable sheets}
\author{M. Bordag\footnote{bordag@itp.uni-leipzig.de}}
\affiliation{ Leipzig University, Institute for Theoretical Physics,  04109 Leipzig, Germany}


%
\begin{abstract}
We consider a flat lattice of dipoles modeled by harmonic oscillators interacting with the \elm field in dipole approximation. Eliminating the variables from the coupled equations of motion, we come to effective Maxwell equations. These allow for taking the lattice spacing $a$ to zero. As a result, we obtain reflection coefficients for the scattering of \elm waves off the sheet. These are a generalization of that known from the hydrodynamic model. For instance, we get a non trivial scattering for polarizability perpendicular to the sheet. Also we show that the case of a sheet polarizable parallel to the sheet, can be obtained in a natural way from a plasma layer of finite thickness. As an alternative approach we discuss the elimination of the \elm fields resulting in effective equations for the oscillators. These shown, for $a\to 0$, divergent behavior, resulting from the electrostatic   interaction of the dipoles.
\end{abstract}
%
%
\maketitle
\setcounter{page}{0}
\section{Introduction}
Monoatomically thin polarizable sheets are interesting as idealizations and as physical models as well. For the latter one may think of   graphene or C$_{60}$ and of thin sheets whose perpendicular degrees of freedom are frozen. Idealization starts with 'hard' boundary conditions, as describing an ideally conducting surface, or membranes. Softening of boundaries results in 'semitransparent' boundaries, e.g., described by delta function potentials. It must be mentioned, that an infinitesimal thin sheet cannot be obtained from shrinking the thickness of a dielectric slab with finite permeability to zero.

In general, interest is such sheets dates back at least to \cite{fett73-81-367}, considering 2-dimensional electron gas. Recently, interest renewed in \cite{para12-89-085021,milt13-36-193} and in \cite{bart13-15-063028}. One of the questions discussed  is the following. In macroscopic electrodynamics, a piece of matter enters trough its permeability $\ep(\x)$. Deforming the piece into a flat sheet, its responds disappears for any finite $\ep(\x)$. This prompts for a description using a delta function, $\ep(\x)=1+\lambda \delta(z)$, as done in \cite{para12-89-085021} for a sheet in the $x,y$-plane. Another approach was taken in \cite{bart13-15-063028}, who models a sheet in 2 ways, by a 2-dimensional lattice of dipoles and by an amorphous sheet of dipoles starting from a non-relativistic treatment. One question is what is the response of such sheet to an electromagnetic field.

The response of a thin sheet, including calculation of  cohesive energy and Casimir effect, was investigated in detail in  \cite{BIII,BV}  within the hydrodynamic model. This model, by construction, allows for polarizability parallel to the sheet only. Perpendicular polarization was discussed in the recent \cite{bart13-15-063028} and applied to the van der Waals interaction between an atom and the sheet as well as in \cite{bart14}, where it was applied to the scattering off the sheet. Perpendicular polarizability was also discussed in \cite{para12-89-085021} with the conclusion that it does not produce any effect (see the paragraph just before section III.A. there).

Actual interest in the response of thin sheets comes also from graphene with electronic excitations described by a Dirac equation. There is a vast literature on this topic and we mention only the Casimir effect calculated within this model in \cite{bord09-80-245406} and the recent investigation of surface plasmons \cite{bord14-89-035421} (and literature cited therein).

In the present paper we take an approach starting from a physical model for the sheet consisting of dipoles on a plane   lattice. These are realized as charged mass points allowed for harmonic motion around a   charge of opposite sign resting at the corresponding lattice point. We take a simple 2-dimensional square lattice in the plane $z=0$. We write down the classical action for these point charges and the electromagnetic field and make the dipole approximation in this action. The resulting equations of motion are the Maxwell equations with source consisting of the point charges and Newton's equations for the motion of the mass points in an electromagnetic field.  This is a completely standard procedure and the dipole approximation is the only step of approximation.
With these equations, we take the lattice spacing to zero and further we investigate the resulting equations.

To proceed, there are at least 2 ways. First, one solves the equations for the mass points and inserts these into the Maxwell equations thus eliminating the dipoles. The resulting effective equations are not easy to solve because of the lattice. However, these equations allow for the limit of shrinking the lattice spacing to zero. After that, the polarization of the sheet is represented by delta functions and their derivatives and the equations can be solved using well known methods. For instance, we calculate the resulting reflection coefficients for the scattering off the sheet and obtain a generalization of the hydrodynamic model. Also, we get non trivial scattering for a sheet polarizable perpendicularly.

A second way to proceed solves the Maxwell equations with the dipoles as sources, which is trivial since in free space, and inserts the solutions into the equations of motion for the dipoles. This is the way used in \cite{bart13-15-063028}. Here one hits the problem of electrostatic self energy of point charges and the  divergence for vanishing lattice spacing, which is specific for a 2-dimensional sheet. In this way, one can investigate, for instance, the excitations of the dipoles.

As for the Casimir effect for 2 parallel sheets, one can calculate it from either the electromagnetic excitations proceeding the first way, or from the excitation of the dipoles, proceeding the second way. For the hydrodynamic model, these two ways were shown in \cite{bord07-76-065011} to give the same result. Here we discuss that this equivalence must hold for perpendicular polarization too.

The paper is organized as follows. In the next section we collect the known formulas for the electromagnetic field and point charges. In section III we derive the equations of motion and make the dipole approximation. In one subsection we consider in-plane polarizability and calculate the reflection coefficients. in the other subsection we consider the equations for the dipoles. In section IV we discuss the equations for the electronic excitations. Further go the conclusions and  technical appendix.

Throughout the paper we use unrationalized Gaussian units.

\section{The model and notations}
We have to work with vectors in 3-dimensional space and on the 2-dimensional sheet at $z=0$. We use the notations
\be \x=\left({\s\atop z}\right),\quad \s=\left({x\atop y}\right),
\label{1.1}\ee
and, for the lengths of these   vectors,
\be x=|\x|=\sqrt{s^2+z^2},\quad s=|\s|.
\label{1.2}\ee
As a convention, we put all  vectors in bold and denote their lengths' by the same letter, taken not bold.
The lattice is given by
\be \s_n=a\n,\quad \n=\left({n_1\atop n_2}\right).
\label{1.3}\ee
Wherever a vector $\s$ or $\s_n$ appears multiplied by a 3-dimensional vector,
\be     \s \to \left({\s\atop 0}\right),\quad  \s_n \to \left({\s_n\atop 0}\right),
\label{1.4}\ee
is assumed.

The action of the system,
\be S=\int d^4x \ {\cal L}=S_{\rm matter}+S_{\rm int}+S_{\rm ED},
\label{1.5}\ee
consists of the usual electrodynamic part,
\be {\cal L}_{\rm ED}=-\frac{1}{16\pi}F_{\mu\nu}^2=\frac{1}{8\pi}(E^2-B^2),
\label{1.6}\ee
with $A^\mu=(\Phi,\A)$ and
\be \E=-\nabla\Phi-\frac1c \pa_t\A,\quad \B={\rm rot} \A,
\label{1.7}\ee
the interaction,
\be     {\cal L}_{\rm int}=-\frac{1}{c}j_\mu A^\mu
        =  -\rho\Phi +\frac{1}{c}\j\A ,
\label{1.8}\ee
and the matter part,
\be     {\cal L}_{\rm matter}
        =\sum_{\n}\frac{m}{2}\left(\dot{\bxi}_n^2- \Omega^2\bxi_n^2\right),
\label{1.9}\ee
constituting charged 3-dimensional harmonic oscillators with mass $m$ at each lattice site. This is different from the matter part in the hydrodynamic model, where a continuous medium is taken from the very beginning and there is no restoring force. As written in \Ref{1.9}, the oscillators are isotope, but can easily be generalized by taking for each spatial  direction its own $\Om$. Let us mention, that the Lagrangian \Ref{1.9} covers the simplest models for atomic polarizability as well as for displacement polarizability (see, e.g., chapt.27 in \cite{Ashcroft}).

Multiplying the displacement $\bxi$  by the charge  we get the dipole moment,
\be \p_\n=e\bxi_\n,
\label{1.91}\ee
at site $\n$. We mention that both, $\bxi_\n$ and $\p_\n$, are functions of $t$, which we do not indicate explicitly.

The model of   charged oscillators produces a charge density from point charges,
\be     \rho(\x)=e\sum_{\n}\left(\delta^3(x-(\s_{\n}+\bxi_{\n}))
                    -\delta^3(x-\s_{\n})\right),
\label{1.10}\ee
with displacement $\xi_\n$ around a lattice site $\s_\n$. Here also the neutralizing charges are included. The corresponding current is
\be     \j(\x)=e\sum_{\n}\dot{\bxi}_{\n}\, \delta^3(x-(\s_{\n}+\bxi_{\n})),
\label{1.11}\ee
where the dot, as in \Ref{1.9}, denotes the time derivative. With these notations, the interaction part of the action takes the form
\bea S_{\rm int} &=&
         \int d^4x \left(-\rho(\x)\Phi(\x)+\frac{1}{c}\j(\x)\A(\x)\right),
\nn\\ &=&
          e\int dx_0 \sum_{\n}
            \Big[-\left(\Phi(\s_n+\bxi_{\n})-\Phi(\s_{\n})\right)
     \nn\\ &&
                   +\frac{1}{c}\dot{\bxi}_{\n}\A(\s_{\n}+\bxi_{\n})\Big].
\label{1.12}\eea
In this way, the system, consisting of point charges and of the electromagnetic field, is specified. Of course, these formulas are in no way new, however these underline, that so far no approximation is made.

Next we do the dipole approximation. It amounts to an expansion of the interaction part up to first order in the elongations $\bxi_\n$. We get from \Ref{1.12}
\be S_{\rm int}^{\rm dipole}=
       e\int dx_0 \sum_{\n}
            \left[-\bxi_\n\nabla\Phi(\s_\n)+\frac{1}{c}\dot{\bxi}_\n
            A(\s_\n)\right].
\label{1.13}\ee
and, using \Ref{1.7},
\be S_{\rm int}^{\rm dipole}=
      e\int dx_0 \sum_{\n}
            \left[\bxi_\n\E(\s_\n)+\frac1c\pa_t(\bxi_\n\A(\s_\n)) \right].
\label{1.14}\ee
The last term is a total derivative and does not enter the equations of motion. So we drop it. With \Ref{1.91} we get
\be      S_{\rm int}^{\rm dipole}=       \int dx_0 \sum_{\n}  \p_\n\,\E(\s_\n),
\label{1.15}\ee
which is the usual interaction of a dipole with the electric field.
The dipole approximation can be done also in the charge density \Ref{1.10} and in the current \Ref{1.11},
\bea \rho^{\rm dipole}(\x) &=& -e\sum_\n \bxi_\n\nabla\delta^3(\x-\s_\n),
\nn\\       \j^{\rm dipole}(\x) &=& e\sum_\n \dot{\bxi}_\n \delta^3(\x-\s_\n),
\label{1.14a}\eea
where the gradient differentiates with respect to $x$.

The equations of motion, which can be derived in a standard way from the above action, are
the Maxwell equations,
\be\begin{array}{rclrcl}{\rm rot}\E(\x)+\frac1c\pa_t\B(\x)&=&0,
    &{\rm div}\E(\x)&=&4\pi\rho,
\\[4pt]  {\rm rot}\B(\x)-\frac1c\pa_t\E(\x)&=&\frac{4\pi}{c}\,\j(\x),\ \
    & {\rm div}\B(\x)&=&0,
\end{array}
\label{1.16}\ee
and the equations of motion for the oscillators,
\be m\left(\pa_t^2+\Omega^2\right)\bxi_\n=  e\E(\s_\n),
\label{1.17}\ee
at each lattice site.

For the following, it is convenient to rewrite the Maxwell equations by eliminating the magnetic   field. Standard manipulations give
\bea    {\rm div}\E&=&4\pi\rho,    \nn\\
        \left(-\frac{1}{c^2} \pa_t^2+\Delta-\nabla\circ\nabla\right)\E
                    &=& \frac{4\pi}{c^2}\pa_t\j.
\label{1.18}\eea
Using Gauss's law, the second line    can also be written in the form
\be      \left(-\frac{1}{c^2} \pa_t^2 +\Delta\right) \E=
        {4\pi}  \left(\nabla \rho+\frac{1}{c^2}\pa_t \j\right).
\label{1.19}\ee
For completeness we mention also the equation for the magnetic field,
\be \left(-\frac{1}{c^2} \pa_t^2 +\Delta\right) \B=-\frac{4\pi}{c}{\rm rot}\j,
\label{1.20}\ee
which we  do not need in the following.

\section{Equations for the \elm field}
In order to proceed, two ways are possible. The first is to solve the equations \Ref{1.17} for the oscillators and to insert the solution into the equations \Ref{1.18} or \Ref{1.19} for the field strengths. This gives the equations including the response of the oscillators and allow, for instance, to calculate the reflection coefficients for the scattering of electromagnetic waves off the sheet. The second way is to proceed in the reverse order, first solving the equations for the electromagnetic field with sources, and then to insert these solutions into the equation \Ref{1.17} for the oscillators. This would allow to investigate the excitations in the sheet. This is discussed in the next section.

Here we proceed in the first way. We consider equation \Ref{1.19} and insert the sources in dipole approximation from \Ref{1.14a},
\bea &&\left(-\frac{1}{c^2}\pa_t^2+\Delta\right)\E(\x)
\nn\\&&=4\pi e \sum_n \left(-\nabla\circ\nabla+\frac{1}{c^2}\pa_t^2\right)\bxi_\n\delta^3(\x-\s_\n).
\label{2.1}\eea
In the right side, the spatial derivatives act only on the delta function, whereas the time derivative acts only on $\bxi_\n$.

The equation \Ref{1.17} for the displacement  can be solved easily using Fourier transform in the time variable, or by assuming harmonic time dependence $\sim\exp(-i\om t)$ everywhere,
\be \bxi_\n=\frac{  e}{m(-\om^2 +\Om^2)}\,\E(\s_\n).
\label{2.2}\ee
Then the dipole moment \Ref{1.91} is
\be \p_\n= \frac{\al(\om)}{4\pi}\E(\s_\n)
\label{2.3}\ee
with
\be \al(\om)=\frac{4\pi e^2}{m \Om^2 (1-\om^2/\Om^2)}
\label{2.4}\ee
and $\al(0)$ is the static polarizability.
In this way, inserting \Ref{2.3} into \Ref{2.1}, we get
\bea &&\left(\frac{\om^2}{c^2}+\Delta\right)\E(\x)
\\\nn&&=- \al(\om)\sum_\n\E(\s_n)
\left(\nabla\circ\nabla+\frac{\om^2}{c^2} \right) \delta^3(\x-\s_\n),
\label{2.5}\eea
where also for $\E(\x)$ harmonic time dependence is assumed. We mention that, according to the convention \Ref{1.4}, the electric field in the right side is to be taken at $z=0$.

Next we take the continuums limit. That is, we assume the lattice spacing $a\to0$. The lattice sum turns into the corresponding 2-dimensional integration and the lattice variables \Ref{1.3} become continuous,
\bea        a^2\sum_\n & \to & \int d^2s' ,\nn\\
            s_\n & \to & s'.
\label{2.6}\eea
Accordingly, we have $\bxi_\n\to \bxi(\s)$ and 
\be n= \frac{1}{a^2}
\label{2.7}\ee
is the density per unit area. Doing this limit in equation \Ref{2.5} we come to
\bea&& \left(\frac{\om^2}{c^2}+\Delta\right)\E(\x)
\\\nn&&=
- \al(\om)n\int d^2s'\, \E(\s')
\left(\nabla\circ\nabla+\frac{\om^2}{c^2} \right) \delta^3(\x-\s'),
\label{2.6a}\eea
where $\s'$ is taken according to \Ref{1.4}. Carrying out the $s'$-integration and using the notation \Ref{1.1} for $\x$, we get
\be \left(\frac{\om^2}{c^2}+\Delta\right)\E(\x)=
-  n\al(\om)
\left(\nabla\circ\nabla+\frac{\om^2}{c^2} \right) \E(\s)\delta(z).
\label{2.8}\ee
Again, we mention that $\E(\s)$ in the right side does not depend on $z$ so that $z$-derivatives from the gradients act on the delta function only.

Next we split these equations into   parts, one parallel to the plane and the other orthogonal to the plane, using notations
\be     \E=\left({\E_{||} \atop E_3}\right),\quad
        \nabla=\left({\nabla_{||} \atop \pa_z}\right).
\label{2.9}\ee
With these, eqs. \Ref{2.8} can be rewritten in the form
\begin{widetext}%
\bea    \left(\frac{\om^2}{c^2}+\Delta\right)\E_{||}(\x) &=&
        -  n\al(\om)
    \left[\left(\nabla_{||}\circ\nabla_{||}+\frac{\om^2}{c^2} \right) \E_{||}(\s)\delta(z)
        +\nabla_{||}E_3(\s)\delta'(z)\right],
\nn\\
           \left(\frac{\om^2}{c^2}+\Delta\right)  E_3(\x) &=&  -   n\al(\om)
    \left[\nabla_{||}\E_{||}(\s)\delta'(z)+
    E_3(\s)\left(\pa_z^2+\frac{\om^2}{c^2}\right)\delta(z)\right].
\label{2.10}\eea
\end{widetext}
As can be seen, first and second order derivatives of the delta function appear. Moreover, these equations do not separate, at least not in an immediate way. For this reason we consider separately oscillators polarizable in-plane only and perpendicular-to-plane only. An alternative way would consider the polarizability $\al$ in Eq.\Ref{2.3} as diagonal matrix.

\subsection{In-plane polarizability}
We go back to equation \Ref{1.15} and allow in the right side for vectors $\p_{||}$ and $\E_{||}$ only. Going through the subsequent formulas we see  that in the right sides $E_3$ does not appear. Thus we get from \Ref{2.10}
\bea    \left(\frac{\om^2}{c^2}+\Delta\right)\E_{||}(\x) &=&
       -   n\al(\om)
 \nn\\&&    \left(\nabla_{||}\circ\nabla_{||}+\frac{\om^2}{c^2} \right) \E_{||}(\s)
        ]\delta(z),
\nn\\
           \left(\frac{\om^2}{c^2}+\Delta\right)  E_3(\x) &=& -    n\al(\om)
    \nabla_{||}\E_{||}(\s)\delta'(z) .
\label{2.11}\eea
These equations have a triangular structure. The first one contains $\E_{||}$ only and it can be solved on its own. The second equation has $E_3$ in the left side only, which can be calculated once $\E_{||}$ is known.

In order to solve these equations, we make a Fourier transform in the directions parallel to the plane and define
\be \tilde{\E}(\k,z)=\int d^2 s  \ \E(\x)\, e^{-i \k \s}
\label{2.12}\ee
and similar for other quantities. Equations \Ref{2.11} turn into
\bea   \label{2.13}  \left(p^2+\pa_z^2\right)\tilde{\E}_{||}(\k,z) &=&
        -  n \al(\om)
\\\nn&& \left(-\k_{||}\circ\k_{||}+\frac{\om^2}{c^2}\right)
            \tilde{\E}_{||}(\k,0)\delta(z),
\\ \nn       \left(p^2+\pa_z^2\right)\tilde{E}_{3}(\k,z) &=&
        -  n \al(\om) \left(-i\k_{||} \right)
            \tilde{\E}_{||}(\k,0)\delta'(z),
\eea
with the definition
\be p=\sqrt{\frac{\om^2}{c^2}-k^2}
\label{2.p}\ee
for the momentum $p$ in the left sides.

From the last line in \Ref{2.13} the normal component of the electric field can be calculated once the parallel components are known.  The equation for these, i.e., the upper line in \Ref{2.13}, is   two-component. It can be diagonalized by the standard TE and TM polarizations. These are
\be \tilde{\E}^{\rm TE}=\left(\begin{array}{c}-k_2\\k_1\\0\end{array}\right)a^{\rm TE},
\quad
\tilde{\E}^{\rm TM}=\left(\begin{array}{c}k_1i\pa_z\\k_2i\pa_z\\-(k_1^2+k_2^2)\end{array}\right)a^{\rm TM},
\label{2.14}\ee
but we need only the properties
\be     \k\,\tilde{\E}_{||}^{\rm TE}=0,\qquad
    \k \, \circ\k \tilde{\E_{||}}^{\rm TM}=k^2\tilde{\E}_{||}^{\rm TM}.
\label{2.15}\ee
Then the equations read
\bea    \left(p^2+\pa_z^2\right)\tilde{\E}_{||}^{\rm TE}(\k,z) &=&
       -   n\al(\om)
     \frac{\om^2}{c^2}   \tilde{\E}_{||}^{\rm TE}(\k,0)
        \delta(z),
\nn\\
           \left(p^2+\pa_z^2\right) \tilde{\E}_{||}^{\rm TM}(\k,z) &=&   -  n\al(\om)p^2
    \tilde{\E}_{||}^{\rm TM}(\k,0)\delta(z) ,
\label{2.16}\eea
where \Ref{2.p} was used in the right side.

These are typical Schr\"odinger equations with a delta function potential. By well known formulas, the delta function can be rewritten as matching conditions. We display these in Appendix A. From \Ref{A.s1} with $\mu\to  -  n \al(\om) \frac{\om^2}{c^2}$ and $\lambda=0$ we get for the TE polarization
\bea {\rm discont~} \tilde{\E}^{\rm TE}_{||}(\k,z) &=& 0,  \nn\\
        {\rm discont~} \pa_z\tilde{\E}_{||}^{\rm TE}(\k,z) &=& -  n\al(\om)\frac{\om^2}{c^2}
                \tilde{\E}_{||}^{\rm TE}(\k,0),
\label{2.17}\eea
and with
$\mu\to  - n \al(\om)p^2$ and $\lambda=0$ we get for the TM polarization
\bea {\rm discont~} \tilde{\E}^{\rm TM}_{||}(\k,z) &=& 0,  \nn\\
        {\rm discont~} \pa_z\tilde{\E}_{||}^{\rm TM}(\k,z) &=&-   n\al(\om)p^2
                \tilde{\E}_{||}^{\rm TM}(\k,0).
\label{2.18}\eea
The equation outside $z=0$ is just the free wave equation. The scattering solutions are like  \Ref{B.1} with $p$, given by eq.\Ref{2.p}, which is the momentum perpendicular to the plane.

From the matching condition \Ref{2.17} and \Ref{2.18}, the reflection coefficients can be written down using the formulas collected in Appendix B. Using \Ref{B.2} we get
\bea r^{\rm TE}  &=&\frac{-1}{1+i\frac{2p c^2}{  n \al(\om)\om^2 }},\nn\\
    r^{\rm TM}  &=&\frac{-1}{1+i\frac{2}{  n \al(\om)p}}.
\label{2.19}\eea
With these formulas, the problem for in-plane polarizable dipoles is solved.

In fact, the above solution is a generalization of the reflection coefficients obtained from the hydrodynamical model in \cite{BV}. To see the equivalence, we first rewrite \Ref{2.19} using \Ref{2.4},
\bea r^{\rm TE}  &=&\frac{-1}
{1-i\frac{p mc^2}{2\pi n  e^2}\left(1-\frac{\Om^2}{\om^2}\right)},\nn\\
    r^{\rm TM}  &=&\frac{-1}{1-i\frac{m\om^2}
    {2\pi  n e^2p}\left(1-\frac{\Om^2}{\om^2}\right)  }.
\label{2.20}\eea
These equation must be compered with equations (2.14) and (2.15) in \cite{BV} using the notation $\q=2\pi n e^2/(mc^2)$ introduced there in eq. (2.5). It is seen  that these expressions coincide if putting $\Om=0$  in \Ref{2.20}. This is because in the hydrodynamic model no restoring force was assumed. In this way, the model of oscillating point charges, used here, reproduces in the continuums limit the known  reflection coefficients of the hydrodynamic model.

\subsection{Perpendicular polarizability}
We go back to equation \Ref{1.15} and allow for $\p_3$ and $\E_3$ only. As a consequence, in the right hands sides in the equations in Section III only the normal component $E_3$ of the electric field appears. Thus we get from \Ref{2.10}
\bea    \left(\frac{\om^2}{c^2}+\Delta\right)\E_{||}(\x) &=&
        -  n\al(\om)
     \nabla_{||}  E_{3}(\s)
         \delta'(z),
\label{2.21}\\\nn
           \left(\frac{\om^2}{c^2}+\Delta\right)  E_3(\x) &=&  -   n\al(\om)
   E_{3}(\s)\left(\pa_z^2+\frac{\om^2}{c^2}\right)\delta(z) .
\eea
These equations have, like that in the preceding subsection, a triangular structure. The lower line is an equation for $E_3$ alone and the upper line allows to calculate $E_{||}$ from a known $E_3$. Now we have to solve the equation for $E_3$. Since this is one equation only, we have only one polarization at work. Looking at \Ref{2.14} it is clear that this should be the TM polarization since it has a $z$-component.

We apply the Fourier transform \Ref{2.12} to the second line in \Ref{2.21} and get
\be (p^2+\pa_z^2)\tilde{E}_3(\k,z)=
- n\al(\om)\tilde{E}_3(\k,0)\left(\pa_z^2+\frac{\om^2}{c^2}\right)\delta(z).
\label{2.22}\ee
This equation is considered in Appendix A. The matching conditions are given by Eq. \Ref{A.s1} with $\mu\to - n\al(\om) \frac{\om^2}{c^2}$ and $\lambda\to\frac{c^2}{\om^2}$,
\bea    {\rm discont~} \tilde{E}_3(\k,0) &=&0, \nn\\
        {\rm discont~}\pa_z\tilde{E}_3(\k,0) &=& - n\al(\om) k^2\tilde{E}_3(\k,0).
\label{2.23}\eea
We mention that the solution for $\tilde{E}_3(\k,z)$ has a delta function contribution like the last term in \Ref{A.4}, making the right side of Eq.\Ref{2.23} ill defined as it stands. However, let us remember that the argument $z=0$ in $\tilde{\E}(\k,z)$ results from the dipole approximation done in eq. \Ref{1.13}. Therefore, in fact we have to take the limit $z\to0$. In that case the problem does not appear as discussed in the Appendix A.

Using equation \Ref{B.1} with the above mentioned substitution for $\mu$ and $\lambda$, we can write down the reflection coefficient,
\be r_{\rm P}=\frac{-1}{1+i\frac{2p }{  n \al(\om) k^2}},
\label{2.24}\ee
which we gave an index 'P' to denote the case of polarizability perpendicular to the sheet.

\section{Equations for the electronic oscillations}
In this section we follow the second way discussed at the beginning of the preceding section and eliminate the \elm field. Since only the electric field enters Eq.\Ref{1.17}, it is sufficient to invert eq.\Ref{2.1}. This can be done easily using the Green function
\be G_\om(x)=
    \int\frac{d^3\k}{(2\pi)^3}\,\frac{e^{i\k\x}}{\om^2-k^2+i0}
        =-\frac{e^{i\om x}}{4\pi x}
\label{3.1}\ee
(note the convention defined after eq. \Ref{1.2}), where in the last expression one has to understand $\om=\sqrt{\om^2+i0}$. We get from eq. \Ref{2.1}
\bea \E(\x)&=&-4\pi e\sum_\n\int d^3{x'}\,G_\om(\x-\x')
  \nn\\&&   \left(\nabla\circ\nabla+\frac{\om^2}{c^2}\right)\bxi_\n\delta^3(\x'-\s_\n),
\label{3.2}\eea
where the gradients differentiate with respect to $\x'$. Using the Yukawa potential in \Ref{3.1} and \Ref{1.91} we get
\be \E(\x)=4\pi \sum_\n T_\om(x-\s_\n)\p_\n
\label{3.3}\ee
with
\be T_\om(x)=\left(\nabla\circ\nabla+\frac{\om^2}{c^2}\right)\frac{e^{i\om x}}{x}.
\label{3.4}\ee
For $\om=0$, \Ref{3.3} is the static field from the dipoles $\p_\n$.

Now we insert \Ref{3.3} into the equation \Ref{1.17} for the oscillators,
\be m(-\om^2+\Om^2)\bxi_\n=4\pi e  \sum_\m T_\om(s_\n-s_\m+\ep)\p_\m.
\label{3.5}\ee
Here we were forced to introduce some regularization $\ep$ to avoid infinite electrostatic selfenergy of the dipoles. Using \Ref{1.91} and \Ref{2.4} we arrive at
\be \p_\n=\al(\om)\sum_\m T_\om(s_\n-s_\m+\ep)\,\p_\m,
\label{3.6}\ee
which is the equation of motion for the dipoles. The well known problem of the electrostatic self energy can be handled   by dropping the singular contribution from $\m=\n$ in the sum. The other, slightly less ad hoc, treatment is to separate the diagonal contributions, to write them in the left side,
\be (1-\al(\om)T_\om(\ep))\p_\n=\al(\om){\sum_\m}' T_\om(s_\n-s_\m+\ep)\,\p_\m,
\label{3.7}\ee
where the primed sum excludes $\m=\n$, and to remove them by defining with
\be     \frac{\al(\om)}{1-\al(\om)T_\om(\ep)}\to\al(\om)
\label{3.8}\ee
a new coupling constant, considering the original one like an unrenormalized  coupling in quantum field theory.

Either way we arrive at the equation
\be \p_\n=\al(\om) {\sum_\m}'T_\om(\s_\n-\s_m)\,\p_\m.
\label{3.9}\ee
It can be diagonalized by Fourier transform,
\be \tilde{\p}(k)=\sum_\n \p_n e^{-i\k\s_\n}.
\label{3.10}\ee
We get
\be \left(1-\frac{\al(\om)}{a^3}\tilde{T}(a\om,a\k)\right)\tilde{\p}(k)=0
\label{3.11}\ee
with
\be  \tilde{T}(\om,\k)={\sum_n}'T_\om(\n)e^{-i\k\n},
\label{3.12}\ee
where now the prime excludes $\n=0$. Eq.\Ref{3.11} determines the excitation in the sheet and the solutions of
\be \det\left(1-\frac{\al(\om)}{a^3}\tilde{T}(a\om,a\k)\right)=0
\label{3.13}\ee
give their dispersions.

Now we consider small lattice spacing, which is equivalent to $\om a\ll1$ and $\k a\ll1$, i.e., to low frequencies and long wave lengths. At this point the dimensionality of the sheet becomes important. With \Ref{2.7}, the factor
\be \frac{\al(\om)}{a^3}=n\al(\om)\frac{1}{a}
\label{3.14}\ee
in front of $\tilde{T}$ in \Ref{2.5} or \Ref{2.6}, is proportional to $1/a$, thus growing for $a\to0$.

Further we need the behavior of $\tilde{T}(a\om,a\k)$ for $a\to0$. First of all we mention that the effects of retardation do not contribute in leading order. Thus we put $\om=0$ and have to consider
\bea \tilde{T}(0,\k) &=&
{\sum_\n}'\frac{3\n\circ\n-n^2}{n^5}e^{-i\k\n},\nn\\
    &=& \left(\pa_k^2-3\nabla_k\circ\nabla_k\right)J_5(\k),
\label{3.15}\eea
with
\be J_s(\k)={\sum_\n}'\frac{e^{-i\k\n}}{n^s},
\label{3.16}\ee
which is absolute convergent for $\Re(s)>2$. From each derivative one has to increase $s$ by one for this convergence. The expansion for small $\k$ can can be found by Fourier transform and reads
\be J_s(\k)=\frac{\Gamma(\frac{2-s}{2})}{2^s\Gamma(s/2)}k^{s-2}+Z_2(s)-\frac{k^2}{4}Z_2(s-2)+\dots
\label{3.17}\ee
with the Epstein zeta function
\be Z_2(s)={\sum_\n}'\frac{1}{n^s}=\frac{1}{4}\zeta_{\rm R}\left(\frac{s}{2}\right)\beta\left(\frac{s}{2}\right)
\label{3.18}\ee
(note $\n$ is 2-dimensional, $n=|\n|$). The last expression uses $\beta(s)=\sum_{k=0}^\infty\frac{(-1)^k}{(2k+1)^s}$ and was found in \cite{hard19-49-89} (see also \cite{glas73-14-409} or \cite{bart13-15-063028}, section 2.). Taking the derivatives in \Ref{3.15} we get
\be\tilde{T}(\om,\k)={\rm diag}\left(\frac12,\frac12,-1\right)Z_2(3)+O(\om,\k)
\label{3.19}\ee
with $Z_2(3)\simeq 9.0336$. The corrections start with first order in $\om$ and $\k$. The matrix element $\tilde{T}_{33}(0,0)$ coincides with (9) in \cite{bart13-15-063028}.

From the finite result for $\tilde{T}(0,0)$, which can also be inferred directly from the upper line in \Ref{3.15} due to the convergence of the sum, it follows that the limit $a\to0$ makes the Coulomb self interaction in Eq.\Ref{3.11} singular. This is different from a 3-dimensional medium where the factor $1/a$ is absorbed in the density, taken per unit volume in that case.

The physical interpretation is quite obvious. The   spacing $a$ of a two dimensional lattice can in this case not go below the interatomic separation determining the range of validity of the dipole approximation. The spectrum, determined by \Ref{3.13} must be expected to be sensitive to $a$. Again, we mention the difference to the 3-dimensional case where for small $a$ the equation is insensitive to a variation of $a$ allowing for $a=0$ even if this is below the region where the dipole approximation is valid.

From Eq.\Ref{3.13} with the approximation \Ref{3.19}, the spectrum of the excitations has 2 solutions for polarizability parallel to the plane with
\be 1-\frac{\al(\om)n}{2a}Z_2(3)\simeq 0
\label{3.20}\ee
and one for perpendicular with
\be 1+\frac{\al(\om)n}{ a}Z_2(3)\simeq 0.
\label{3.21}\ee
Both do not depend on $\k$ in this approximation. The latter case was also discussed in detail in \cite{bart13-15-063028}, section 2.

\section{Conclusions}
We considered the simplest model for polarizability of a 2-dimensional lattice. Starting from the complete action and making only the dipole approximation, we considered
\begin{enumerate}\item
the effective equations for the \elm field,
\item the effective equations for the dipoles,\end{enumerate}
by eliminating the corresponding variables from the equations of motion. It must be mentioned that this procedure carries over directly to the corresponding quantum theory. One would represent the partition function by an functional integral defined with the considered action and integrates out either the   variables of the polarization or the variables for the \elm field. Such procedure was discussed in detail in \cite{bord07-76-065011}, section 2.

In the first way, it is possible to take the lattice spacing $a\to0$ and to calculate the reflection coefficients for the scattering of the \elm waves off the sheet.
We collect here the results from Section III written in terms of the scattering phase shifts defined in Appendix B,
\bea       \tan \eta_{\rm TE}&=&\frac{n \alpha(\om)}{2}\,\frac{\om^2}{p c^2},
\nn\\        \tan \eta_{\rm TM}&=&\frac{n \alpha(\om)}{2}\,p,
\nn\\        \tan \eta_{\rm P}&=&\frac{n \alpha(\om)}{2}\,\frac{k^2}{p},
\label{4.eta}\eea
where $\eta_{\rm TE}$ and $\eta_{\rm TM}$ are the phase shifts for the corresponding polarizations in case of polarizability only parallel to the sheet and $\eta_{\rm P}$ is for polarizability only perpendicular to the sheet

For a polarizability only parallel to the sheet, we obtain, for the oscillator eigenfrequency $\Om=0$, the reflection coefficients $r_{\rm TE}$ and $r_{\rm TM}$ known from the hydrodynamic model \cite{BV}. For a polarizability only perpendicular to the sheet, only one polarization can couple to the sheet for parity reasons. Its reflection coefficient is different from the coefficients for the parallel polarizability.

The latter result is different from the findings in \cite{para12-89-085021}. In that paper a permeability
\be \ep=1+\lambda \delta(z)
 \label{4.0}\ee
was considered. Within our approach, we consider from \Ref{1.16}   Gauss's law and insert from \Ref{1.14a},
\be {\rm div} \E(\x)=-4\pi e\sum_\n\bxi_\n\nabla\delta(\x-\s_\n).
\label{4.1}\ee
In this expression we take the limit $a\to0$ and get with \Ref{1.91}
\be {\rm div}\left( \E(\x)+4\pi \p(\s)\delta(z)\right)=0.
\label{4.2}\ee
Further we let $a\to0$ in \Ref{2.3} obtaining $\p(\s)=\al(\om)\E(\s)$ (note the convention \Ref{1.4}) and insert that into \Ref{4.2},
\be {\rm div}\left( \E(\x)+4\pi \al(\om)\delta(z)\E(\s)\right)=0
\label{4.3}\ee
and read off the  permeability
\be \ep-1=4\pi\al(\om)\delta(z)
\label{4.4}\ee
confirming the structure of \Ref{4.0}, used in \cite{para12-89-085021}, within our model. So the starting formulas are the same, but the conclusions concerning the polarizability perpendicular to the sheet are different.

The reflection coefficient $ \eta_{\rm P}$ in \Ref{4.eta}  is also different from the finding in \cite{bart14}. It has a similar form, but the polarizability, which in our formulas is given by \Ref{2.4}, has in \cite{bart14} an additional contribution in the parenthesis in the denominator resulting from the electrostatic selfinteraction of the dipoles.

It is interesting to note how the results for the polarizability parallel to the sheet can be obtained in a natural way as limiting case from a slab of finite thickness. Let the slab be formed from two parallel planes of separation $L$ with a plasma in between, producing a permittivity
\be \ep=1-\frac{\om_p^2}{\om^2}
\label{4.ep}\ee
with the plasma frequency
\be \om_p^2=\frac{4\pi n_3e^2}{m}.
\label{4.omp}\ee
We mention that this permittivity corresponds to an isotropic polarizability of the plasma.
In \Ref{4.omp},  $n_3$ is the density per unit volume of the plasma. The reflection coefficients
$r_{\rm TE}^{(L)}$ and $r_{\rm TM}^{(L)}$, where the superscript '(L)' indicates the finite thickness, for the scattering off the slab are well known and displayed in Appendix C, eq.\Ref{C.r}. Now we consider the limit of making the slab thin. It is well known, that for finite $\ep$ these reflection coefficients vanish for $L\to0$. However, we make a point that the density $n_3$ in \Ref{4.omp} is no longer appropriate and that it is natural  to use
\be n_3=n\frac{1}{L}
\label{4.nL}\ee
instead, where $n$ is the density per unit area used in section III and appearing in the reflection coefficients \Ref{2.19} and \Ref{2.20}. With this relation for the densities, we get for the plasma frequency \Ref{4.omp}
\be \om_p^2=\frac{2\pi ne^2}{m}\ \frac{2}{L}\equiv\q c^2\frac{2}{L},
\label{4.omp1}\ee
which, when inserted into \Ref{4.ep} gives a permittivity, growing $\ep\sim 1/L$ for $L\to0$.
The parameter $\q$ is just that discussed at the end of section III.A.

Now the statement, proven in Appendix C, is that the reflection coefficients
$r_{\rm TE}^{(L)}$ and $r_{\rm TM}^{(L)}$
turn for $L\to0$ into that of the plasma shell model. As mentioned at the end of section III.A, these are given by Eq. \Ref{2.20} with $\Om=0$.

We did not try the case when parallel and perpendicular polarizabilities are present both at the same time. We conclude the discussion of the first way in our approach with a remark on the Casimir effect for two parallel sheets of the considered kind, which are semitransparent and can be represented   in the continuum limit by delta functions. For such planes, the Casimir effect was first calculated in \cite{bord92-25-4483} (and reconsidered recently \cite{cast13-87-105020}), for the hydrodynamic model in \cite{BVI}. Using these methods, especially the Lifshitz formula, it can be easily calculated also for the reflection coefficients \Ref{2.19} and \Ref{2.24}, found in section III.

As for the second way, the equations for the oscillators do not allow for a limit $a \to 0$ because of the interplay of dimensionality and Coulomb interaction between dipoles, as also discussed in detail in \cite{bart13-15-063028}. In view of this, the result of the insensitivity of  the scattering of \elm waves off the sheet to the limit $a\to0$  is somehow counterintuitive. On the other side, let us think of the Casimir effect for two sheets in terms of a mode sum over the spectrum of the electronic excitations. For two sheets one can generalize eq.\Ref{3.11} correspondingly. Than the separation $L$ between the sheets would enter $T(\om)$ in \Ref{3.9} making the electrostatic contributions between the sheets non singular. Further one could imagine that after the subtraction of the selfenergies of the individual sheets, which do not depend on $L$, the sensitivity to small $a$ disappears. Of course, a proof of this conjecture would be helpful.

To conclude this discussion we remind that for the hydrodynamic model the equivalence of both ways was shown   in \cite{bord07-76-065011}, section 2. An extension to the model considered here seems feasible.


\appendix
\section{Delta function potentials and matching conditions}
Here we collect the simple formulas allowing to recast a delta function potential into matching conditions. The procedure is to solve the equation everywhere except for the point, where the delta function is sitting, $z=0$ for simplicity, and to supplement by matching conditions. We adopt the notation
\be {\rm discont~} \Phi(x) =\lim_{\ep\to0}(\Phi(x+\ep)-\Phi(x-\ep))
\label{A.0}\ee
for the discontinuity. The equation reads
\be \left(p^2+\pa_z^2\right)\Phi(z)=\mu \Phi(0)(1+\lambda \pa_z^2)\delta(z) .
\label{A.1}\ee
We make the Ansatz
\be \Phi(z)=\Phi_-(z)\Theta(-z)+\Phi_+(z)\Theta(z)+h\,\delta(z).
\label{A.4}\ee
The second derivative is
\bea  \pa_z^2\Phi(z)&=&(\Phi_+(0)-\Phi_-(0))\delta'(z)+
        (\Phi'_{+}(0)-\Phi'_{-}(0))\delta(z)
        \nn\\&&+h\delta''(z)+\Phi''_-(z)\Theta(-z)+\Phi''_+(z)\Theta(z).
\label{A.5}\eea
From eq. \Ref{A.1}, the equations
\be \left(p^2+\pa_z^2\right)\Phi_{\pm}(z)=0
\label{A.2}\ee
hold for $z\ne0$, i.e., 'outside' the delta function.
Inserting \Ref{A.4} and \Ref{A.5} into equation \Ref{A.1} and matching the delta functions and their derivatives, we get the matching conditions
\bea    {\rm discont~} \Phi(x) &=&0, \nn\\
        {\rm discont~} \Phi'(x) &=& \mu (1-\lambda p^2)\Phi(0),\nn\\
            h  &=& \mu \lambda\Phi(0).
\label{A.s1}\eea
For $\lambda =0$ these are the matching conditions for a delta function potential well known from, e.g., quantum mechanics. These apply to the parallel polarizability in Section III.A.
For $\lambda\ne0$, as it appears in section III.B for the perpendicular polarizability, there is a problem with the delta function in the Ansatz \Ref{A.4} and $\Phi(z)$ at $z=0$ in the right side of the equation \Ref{A.1}. As it stands, it is not well defined. This problem can be avoided only if $\Phi(0)$ in the right side of \Ref{A.1} is to be understood as limit $z\to0$ in $\Phi(z)$, which is then $\Phi_{+}(0)$, or equivalently, $\Phi_{-}(0)$.\\

\section{Reflection coefficients}
We consider the one dimensional scattering setup with a function
\be \Phi(z)=\left(e^{ipz}+r\,e^{-ipz}\right)\Theta(-z)+t\,e^{ipz}\Theta(z)
\label{B.1}\ee
with reflection coefficient $r$ and transmission coefficient $t$. For the matching conditions \Ref{A.s1} these are
\be r=\frac{-1}{1-\frac{2ip}{\mu(1-\lambda p^2)}},\quad t=\frac{ 1}{1-\frac{\mu(1-\lambda p^2)}{2ip}},
\label{B.2}\ee
where $\mu$ and $\lambda$ are defined in Eq. \Ref{A.1}.
We mention that these coefficients can also be written in terms of phase shifts (see, e.g., Eq.(2.14) in \cite{BV}),
\bea r&=&i \sin(\eta)e^{i\eta}=\frac{-1}{1+i\cot\eta}, \nn\\
 t&=&  \cos(\eta)e^{i\eta}=\frac{1}{1-i \tan\eta},
\eea
with
\be     \eta=-\arctan\frac{\mu(1-\lambda p^2)}{p}.
\ee

\section{Slab of finite thickness}
Here we collect the formulas for scattering off a slab of finite thickness $L$ filled with a plasma having permittivity $\ep$. Let the surface of the slab be formed by two planes, intersection the $z$-axis in $z=0$ and in $z=L$. For the momenta $\k$ parallel to the surfaces and $p$ perpendicular to the surfaces we take the same notations as in section III.
In addition, here we have also a perpendicular momentum, $q$, in between the surfaces.
These momenta obey the relations
\bea \om^2&=&c^2(\k^2+p^2),
\nn\\ \ep\om^2&=&c^2(\k^2+q^2),
\label{C.1}\eea
following from the wave equations outside and inside the slab.

With these notations, the reflection coefficients read
\bea r_{\rm TE}^{\rm (L)}&=& \frac{\frac{q-p}{q+p}\left(e^{2iqL}-1\right)}
                        {  1-\left(\frac{q-p}{q+p}\right)^2e^{2iqL}},
\nn\\
 r_{\rm TM}^{\rm (L)}&=& \frac{\frac{q-\ep p}{q+\ep p}\left(e^{2iqL}-1\right)}
                        {  1-\left(\frac{q-\ep p}{q+\ep p}\right)^2e^{2iqL}}.
\label{C.r}\eea
One should note that frequently only the denominators enter in applications like Lifshitz formula.

Now the statement is that with
\be     \ep=1-\frac{2\q c^2}{\om^2 L}
\label{C.ep}\ee
and \Ref{C.1}, the relations
\be  r_{\rm TE}^{\rm (L)}=\frac{-1}{1-i\frac{p}{\q}}+O(L),  \quad
         r_{\rm TM}^{\rm (L)}=\frac{-1}{1-i\frac{\om^2}{pc^2\q}}+O(\sqrt{L}),
\label{C.3}\ee
hold for $L\to0$. These can be verified by inserting \Ref{C.ep} into \Ref{C.r} after some calculation. Compensations between numerators and denominators appear.
The leading terms in \Ref{C.3} are just the reflection coefficients for the plasma shell model, Eqs.\Ref{2.14} and \Ref{2.15} in \cite{BV}


\begin{thebibliography}{16}%
\makeatletter
\providecommand \@ifxundefined [1]{%
 \@ifx{#1\undefined}
}%
\providecommand \@ifnum [1]{%
 \ifnum #1\expandafter \@firstoftwo
 \else \expandafter \@secondoftwo
 \fi
}%
\providecommand \@ifx [1]{%
 \ifx #1\expandafter \@firstoftwo
 \else \expandafter \@secondoftwo
 \fi
}%
\providecommand \natexlab [1]{#1}%
\providecommand \enquote  [1]{``#1''}%
\providecommand \bibnamefont  [1]{#1}%
\providecommand \bibfnamefont [1]{#1}%
\providecommand \citenamefont [1]{#1}%
\providecommand \href@noop [0]{\@secondoftwo}%
\providecommand \href [0]{\begingroup \@sanitize@url \@href}%
\providecommand \@href[1]{\@@startlink{#1}\@@href}%
\providecommand \@@href[1]{\endgroup#1\@@endlink}%
\providecommand \@sanitize@url [0]{\catcode `\\12\catcode `\$12\catcode
  `\&12\catcode `\#12\catcode `\^12\catcode `\_12\catcode `\%12\relax}%
\providecommand \@@startlink[1]{}%
\providecommand \@@endlink[0]{}%
\providecommand \url  [0]{\begingroup\@sanitize@url \@url }%
\providecommand \@url [1]{\endgroup\@href {#1}{\urlprefix }}%
\providecommand \urlprefix  [0]{URL }%
\providecommand \Eprint [0]{\href }%
\providecommand \doibase [0]{http://dx.doi.org/}%
\providecommand \selectlanguage [0]{\@gobble}%
\providecommand \bibinfo  [0]{\@secondoftwo}%
\providecommand \bibfield  [0]{\@secondoftwo}%
\providecommand \translation [1]{[#1]}%
\providecommand \BibitemOpen [0]{}%
\providecommand \bibitemStop [0]{}%
\providecommand \bibitemNoStop [0]{.\EOS\space}%
\providecommand \EOS [0]{\spacefactor3000\relax}%
\providecommand \BibitemShut  [1]{\csname bibitem#1\endcsname}%
\let\auto@bib@innerbib\@empty
\bibitem [{\citenamefont {Fetter}(1973)}]{fett73-81-367}%
  \BibitemOpen
  \bibfield  {author} {\bibinfo {author} {\bibfnamefont {A.~L.}\ \bibnamefont
  {Fetter}},\ }\bibfield  {title} {\enquote {\bibinfo {title} {{Electrodynamics
  of a Layered Electron-Gas.1. Single Layer}},}\ }\href@noop {} {\bibfield
  {journal} {\bibinfo  {journal} {Ann.~Phys.}\ }\textbf {\bibinfo {volume}
  {81}},\ \bibinfo {pages} {367--393} (\bibinfo {year} {1973})}\BibitemShut
  {NoStop}%
\bibitem [{\citenamefont {Parashar}\ \emph {et~al.}(2012)\citenamefont
  {Parashar}, \citenamefont {Milton}, \citenamefont {Shajesh},\ and\
  \citenamefont {Schaden}}]{para12-89-085021}%
  \BibitemOpen
  \bibfield  {author} {\bibinfo {author} {\bibfnamefont {Prachi}\ \bibnamefont
  {Parashar}}, \bibinfo {author} {\bibfnamefont {Kimball~A.}\ \bibnamefont
  {Milton}}, \bibinfo {author} {\bibfnamefont {K.~V.}\ \bibnamefont {Shajesh}},
  \ and\ \bibinfo {author} {\bibfnamefont {M.}~\bibnamefont {Schaden}},\
  }\bibfield  {title} {\enquote {\bibinfo {title} {Casimir interaction energy
  for magneto-electric $\delta$-function plates},}\ }\href@noop {} {\bibfield
  {journal} {\bibinfo  {journal} {Phys.~Rev.~D}\ }\textbf {\bibinfo {volume}
  {86}},\ \bibinfo {pages} {085021} (\bibinfo {year} {2012})}\BibitemShut
  {NoStop}%
\bibitem [{\citenamefont {Milton}\ \emph {et~al.}(2013)\citenamefont {Milton},
  \citenamefont {Parashar}, \citenamefont {Schaden},\ and\ \citenamefont
  {Shajesh}}]{milt13-36-193}%
  \BibitemOpen
  \bibfield  {author} {\bibinfo {author} {\bibfnamefont {Kimball~A.}\
  \bibnamefont {Milton}}, \bibinfo {author} {\bibfnamefont {Prachi}\
  \bibnamefont {Parashar}}, \bibinfo {author} {\bibfnamefont {M.}~\bibnamefont
  {Schaden}}, \ and\ \bibinfo {author} {\bibfnamefont {K.~V.}\ \bibnamefont
  {Shajesh}},\ }\bibfield  {title} {\enquote {\bibinfo {title} {Electromagnetic
  semitransparent $\delta$-function plate: Casimir interaction energy between
  parallel infinitesimally thin plates},}\ }\href@noop {} {\bibfield  {journal}
  {\bibinfo  {journal} {Nuovo Cimento}\ }\textbf {\bibinfo {volume} {36}},\
  \bibinfo {pages} {193} (\bibinfo {year} {2013})}\BibitemShut {NoStop}%
\bibitem [{\citenamefont {Barton}(2013)}]{bart13-15-063028}%
  \BibitemOpen
  \bibfield  {author} {\bibinfo {author} {\bibfnamefont {G.}~\bibnamefont
  {Barton}},\ }\bibfield  {title} {\enquote {\bibinfo {title} {Casimir effects
  in monatomically thin insulators polarizable perpendicularly: nonretarded
  approximation},}\ }\href@noop {} {\bibfield  {journal} {\bibinfo  {journal}
  {New~J.~Phys.}\ }\textbf {\bibinfo {volume} {15}},\ \bibinfo {pages} {063028}
  (\bibinfo {year} {2013})}\BibitemShut {NoStop}%
\bibitem [{\citenamefont {Barton}(2004)}]{BIII}%
  \BibitemOpen
  \bibfield  {author} {\bibinfo {author} {\bibfnamefont {G.}~\bibnamefont
  {Barton}},\ }\bibfield  {title} {\enquote {\bibinfo {title} {Casimir energies
  of spherical plasma shells},}\ }\href@noop {} {\bibfield  {journal} {\bibinfo
   {journal} {J. Phys. A: Math. Gen.}\ }\textbf {\bibinfo {volume} {37}},\
  \bibinfo {pages} {1011--1049} (\bibinfo {year} {2004})}\BibitemShut {NoStop}%
\bibitem [{\citenamefont {Barton}(2005{\natexlab{a}})}]{BV}%
  \BibitemOpen
  \bibfield  {author} {\bibinfo {author} {\bibfnamefont {G.}~\bibnamefont
  {Barton}},\ }\bibfield  {title} {\enquote {\bibinfo {title} {{Casimir effects
  for a flat plasma sheet: I. Energies}},}\ }\href@noop {} {\bibfield
  {journal} {\bibinfo  {journal} {J. Phys. A: Math. Gen.}\ }\textbf {\bibinfo
  {volume} {38}},\ \bibinfo {pages} {2997--3019} (\bibinfo {year}
  {2005}{\natexlab{a}})}\BibitemShut {NoStop}%
\bibitem [{\citenamefont {Bordag}\ \emph {et~al.}(2009)\citenamefont {Bordag},
  \citenamefont {Fialkovsky}, \citenamefont {Gitman},\ and\ \citenamefont
  {Vassilevich}}]{bord09-80-245406}%
  \BibitemOpen
  \bibfield  {author} {\bibinfo {author} {\bibfnamefont {M.}~\bibnamefont
  {Bordag}}, \bibinfo {author} {\bibfnamefont {I.~V.}\ \bibnamefont
  {Fialkovsky}}, \bibinfo {author} {\bibfnamefont {D.~M.}\ \bibnamefont
  {Gitman}}, \ and\ \bibinfo {author} {\bibfnamefont {D.~V.}\ \bibnamefont
  {Vassilevich}},\ }\bibfield  {title} {\enquote {\bibinfo {title} {{Casimir
  interaction between a perfect conductor and graphene described by the Dirac
  model}},}\ }\href@noop {} {\bibfield  {journal} {\bibinfo  {journal}
  {Phys.~Rev.~B}\ }\textbf {\bibinfo {volume} {80}},\ \bibinfo {pages} {245406}
  (\bibinfo {year} {2009})}\BibitemShut {NoStop}%
\bibitem [{\citenamefont {Bordag}\ and\ \citenamefont
  {Pirozhenko}(2014)}]{bord14-89-035421}%
  \BibitemOpen
  \bibfield  {author} {\bibinfo {author} {\bibfnamefont {M.}~\bibnamefont
  {Bordag}}\ and\ \bibinfo {author} {\bibfnamefont {I.G.}\ \bibnamefont
  {Pirozhenko}},\ }\bibfield  {title} {\enquote {\bibinfo {title}
  {{Transverse-electric surface plasmon for graphene in the Dirac equation
  model}},}\ }\href@noop {} {\bibfield  {journal} {\bibinfo  {journal}
  {Phys.~Rev.~B}\ }\textbf {\bibinfo {volume} {89}},\ \bibinfo {pages} {035421}
  (\bibinfo {year} {2014})}\BibitemShut {NoStop}%
\bibitem [{\citenamefont {Bordag}(2007)}]{bord07-76-065011}%
  \BibitemOpen
  \bibfield  {author} {\bibinfo {author} {\bibfnamefont {M.}~\bibnamefont
  {Bordag}},\ }\bibfield  {title} {\enquote {\bibinfo {title} {On the
  interaction of a charge with a thin plasma sheet},}\ }\href@noop {}
  {\bibfield  {journal} {\bibinfo  {journal} {Phys.~Rev.~D}\ }\textbf {\bibinfo
  {volume} {76}},\ \bibinfo {pages} {065011} (\bibinfo {year}
  {2007})}\BibitemShut {NoStop}%
\bibitem [{\citenamefont {Ashcroft}\ and\ \citenamefont
  {Mermin}(1976)}]{Ashcroft}%
  \BibitemOpen
  \bibfield  {author} {\bibinfo {author} {\bibfnamefont {N.W.}\ \bibnamefont
  {Ashcroft}}\ and\ \bibinfo {author} {\bibfnamefont {N.D.}\ \bibnamefont
  {Mermin}},\ }\href@noop {} {\emph {\bibinfo {title} {Solid State Physics}}}\
  (\bibinfo  {publisher} {Thomson Learning},\ \bibinfo {address} {United
  States},\ \bibinfo {year} {1976})\BibitemShut {NoStop}%
\bibitem [{\citenamefont {Hardy}(1919)}]{hard19-49-89}%
  \BibitemOpen
  \bibfield  {author} {\bibinfo {author} {\bibfnamefont {G.H.}\ \bibnamefont
  {Hardy}},\ }\href@noop {} {\bibfield  {journal} {\bibinfo  {journal}
  {Mess.Math}\ }\textbf {\bibinfo {volume} {49}},\ \bibinfo {pages} {89}
  (\bibinfo {year} {1919})}\BibitemShut {NoStop}%
\bibitem [{\citenamefont {Glasser}(1973)}]{glas73-14-409}%
  \BibitemOpen
  \bibfield  {author} {\bibinfo {author} {\bibfnamefont {M.L.}\ \bibnamefont
  {Glasser}},\ }\bibfield  {title} {\enquote {\bibinfo {title} {{The evaluation
  of lattice sums. I. Analytic procedures}},}\ }\href@noop {} {\bibfield
  {journal} {\bibinfo  {journal} {J.~Math.~Phys.}\ }\textbf {\bibinfo {volume}
  {14}},\ \bibinfo {pages} {409} (\bibinfo {year} {1973})}\BibitemShut
  {NoStop}%
\bibitem [{\citenamefont {Barton}(2014)}]{bart14}%
  \BibitemOpen
  \bibfield  {author} {\bibinfo {author} {\bibfnamefont {G.}~\bibnamefont
  {Barton}},\ }\href@noop {} {\enquote {\bibinfo {title} {{Monolayers
  Polarizable Perpendicularly: The Maxwellian Response Functions}},}\ }
  (\bibinfo {year} {2014}),\ \bibinfo {note} {private
  communication}\BibitemShut {NoStop}%
\bibitem [{\citenamefont {Bordag}\ \emph {et~al.}(1992)\citenamefont {Bordag},
  \citenamefont {Hennig},\ and\ \citenamefont {Robaschik}}]{bord92-25-4483}%
  \BibitemOpen
  \bibfield  {author} {\bibinfo {author} {\bibfnamefont {M.}~\bibnamefont
  {Bordag}}, \bibinfo {author} {\bibfnamefont {D.}~\bibnamefont {Hennig}}, \
  and\ \bibinfo {author} {\bibfnamefont {D.}~\bibnamefont {Robaschik}},\
  }\bibfield  {title} {\enquote {\bibinfo {title} {{Vacuum energy in quantum
  field theory with external potentials concentrated on planes}},}\ }\href@noop
  {} {\bibfield  {journal} {\bibinfo  {journal} {J.~Phys.~A: Math.~Gen.}\
  }\textbf {\bibinfo {volume} {25}},\ \bibinfo {pages} {4483} (\bibinfo {year}
  {1992})}\BibitemShut {NoStop}%
\bibitem [{\citenamefont {Castaneda}\ \emph {et~al.}({2013})\citenamefont
  {Castaneda}, \citenamefont {Mateos~Guilarte},\ and\ \citenamefont
  {Moreno~Mosquera}}]{cast13-87-105020}%
  \BibitemOpen
  \bibfield  {author} {\bibinfo {author} {\bibfnamefont {J.~M.~Munoz}\
  \bibnamefont {Castaneda}}, \bibinfo {author} {\bibfnamefont {J.}~\bibnamefont
  {Mateos~Guilarte}}, \ and\ \bibinfo {author} {\bibfnamefont {A.}~\bibnamefont
  {Moreno~Mosquera}},\ }\bibfield  {title} {\enquote {\bibinfo {title}
  {{Quantum vacuum energies and Casimir forces between partially transparent
  delta-function plates}},}\ }\href {\doibase {10.1103/PhysRevD.87.105020}}
  {\bibfield  {journal} {\bibinfo  {journal} {Phys.~Rev.~D}\ }\textbf {\bibinfo
  {volume} {{87}}},\ \bibinfo {pages} {{105020}} (\bibinfo {year}
  {{2013}})}\BibitemShut {NoStop}%
\bibitem [{\citenamefont {Barton}(2005{\natexlab{b}})}]{BVI}%
  \BibitemOpen
  \bibfield  {author} {\bibinfo {author} {\bibfnamefont {G.}~\bibnamefont
  {Barton}},\ }\bibfield  {title} {\enquote {\bibinfo {title} {{Casimir effects
  for a flat plasma sheet: II. Fields and stresses}},}\ }\href@noop {}
  {\bibfield  {journal} {\bibinfo  {journal} {J. Phys. A: Math. Gen.}\ }\textbf
  {\bibinfo {volume} {38}},\ \bibinfo {pages} {3021--3044} (\bibinfo {year}
  {2005}{\natexlab{b}})}\BibitemShut {NoStop}%
\end{thebibliography}
%
\end{document}